\documentclass[preprintnumbers,amsmath,amssymb,prc,twocolumn]{revtex4}
\usepackage{graphicx}
\usepackage{amssymb}
\usepackage{amsmath}
\usepackage{ulem}
\usepackage{xcolor,etoolbox}
\usepackage{dcolumn}
\usepackage{bm}
\begin{document}
\title{Astrophysical reaction rates with realistic nuclear level densities}

\author{Sangeeta$^1$, T. Ghosh$^{2,3}$, B. Maheshwari$^{4,5}$, G. Saxena$^{6,*}$, B. K. Agrawal$^{2,3}$}
\affiliation{$^1$Department of Applied Sciences, Chandigarh Engineering College, Landran-140307, India}
\affiliation{$^2$Saha Institute of Nuclear Physics, Kolkata-700064, India}
\affiliation{$^3$Homi Bhabha National Institute, Anushakti Nagar, Mumbai-400094, India}
\affiliation{$^4$Department of Physics, Faculty of Science, University of Zagreb, HR-10000 Zagreb, Croatia}
\affiliation{$^5$Department of Physics, Indian Institute of Technology Ropar, Rupnagar-140001, India}
\affiliation{$^6$Department of Physics (H $\&$ S), Govt. Women Engineering College, Ajmer-305002, India}
\email{gauravphy@gmail.com}

\begin{abstract}
Realistic nuclear level densities (NLDs) obtained within the spectral distribution method (SDM) are employed to study nuclear processes of astrophysical interest. The merit of SDM  lies in the fact that the NLDs corresponding to many body shell model Hamiltonian consisting of residual interaction can be obtained for the full configurational space without recourse to the exact diagnolization of huge matrices.
We  calculate NLDs and s-wave neutron resonance spacings which agree reasonably well with the available experimental data. By employing these NLDs, we compute reaction cross-sections and astrophysical reaction rates for radiative neutron capture in few Fe-group nuclei, and compare them with experimental data as well as with those obtained with NLDs from phenomenological and microscopic mean-field models. The results obtained  for the NLDs from SDM  are able to explain the experimental data quite well. These results are of particular importance since the configuration mixing
through the residual interaction naturally accounts for the collective excitations.  In the  mean-field models,  the collective effects are included through the vibrational and rotational enhancement factors and their NLDs are further normalized at low energies with neutron resonance data.
\end{abstract}
\date{\today}

\maketitle

\section{Introduction}

Neutron capture reactions are critical for a wide variety of applications ranging from medicine \cite{Lima1998}, energy generation \cite{Prelas2016} to nucleosynthesis \cite{Burbidge1957}. Such reactions are fundamentally important by which nearly all of the elements heavier than iron are synthesized through astrophysical s-process or r-process. The astrophysical site where the r-process occurs has been one of the major open questions for last several decades. Supernovae and neutron star mergers are the most favored astrophysical sites for the r-process. Only recently, observations connected to the neutron star merger event GW170817 \cite{Abbott2017} have confirmed the emission of a kilonova afterglow which is powered by the radioactive decay of heavy nuclides produced in the r-process \cite{Arcavi2017,Pian2017}. Efforts are also being made to simulate the neutron capture nucleosynthesis in the laboratory \cite{Wu2021}. This led to a rapid increase in attention on various processes contributing to nuclear reaction network. \par

The extensive radiative neutron capture cross-section data are crucial for complete understanding of r-process nuclear astrophysical network calculations. All such data are not accessible in the accelerator-based experiments and one needs to rely on theoretical estimates. The reaction cross-sections and relevant reaction rates are calculated within a statistical framework \cite{Hauser1952} which primarily requires (i) neutron-nucleus optical model potential (OMP), (ii) $\gamma$-ray strength function ($\gamma$SF), and (iii) nuclear level density (NLD). The uncertainties of $\gamma$SF and NLD  have significant impact on calculated neutron capture rates while the uncertainties due to OMP are relatively smaller \cite{Liddick2016,Dutta2016}. The NLD describes the total number of states accessible in a given nucleus at a specific excitation energy. Various methods have been employed to calculate the NLDs which range from simple phenomenological models based on non-interacting degenerate Fermi gas \cite{Dilg1973,Gilbert1965} to more complex mean-field descriptions \cite{Goriely2008}. In the mean-field approaches, all the collective effects are incorporated in NLDs through the rotational and vibrational enhancement factors. These NLDs are further normalized with the experimental data at the neutron resonances \cite{Goriely2008}. In the shell model, the configuration mixing through the residual interaction naturally accounts for the collective excitations.\par

There are a few approaches to calculate the NLDs within the framework of shell model. One of them is the shell model Monte Carlo (SMMC) \cite{Johnson1992,Lang1993,nakada1997,nakada1998,Nakada2008,Alhassid2015}, which utilizes auxiliary fields to compute the thermal trace for the energy and then NLDs are obtained using inverse Laplace transform. Another approach to calculate the spin and parity dependent shell model level densities is developed using the spectral distribution method (SDM) \cite{Chang1971,French1971,Kota1989,Kota1996,Horoi2003,Horoi2004}. This method allows one to incorporate the many-body effects on the wave functions appropriately and is a basis for the applications of statistical spectroscopy generated by many-body chaos, as modeled by embedded random matrix ensembles, along with the studies on beta decay rates for stellar evolution and supernovae \cite{kar1994,Kota1995,Gomez2011,Kota2018}. Recently, the SDM has been extended for the exact calculation of the first and second Hamiltonian moments for different configurations at fixed spin and parity \cite{Senkov2010,horoi2011,Senkov2013,Senkov2016}. This is a practical tool to construct the NLDs for many body shell model Hamiltonian using full configurational space since it avoids the diagonalization of Hamiltonian matrices of huge dimensions. This method, however, requires an accurate estimation of the shell model ground state energy, which is generally as time consuming as the full shell model calculation. The difficulty has been overcome by using the exponential convergence method \cite{Horoi2002} or the recently developed projected configuration interaction method \cite{Gao2009,Gao20091}. The NLDs obtained from SDM also agree reasonably with the full shell model calculations \cite{Senkov2013}. Recently, extrapolated Lanczos method \cite{Ormand2020} has also been developed for an accurate computation of the level densities described within the configuration space.\par

In the present work, we use realistic NLDs obtained from spectral distribution method followed by an appropriate parity equilibration scheme for $pf$-model space to calculate the neutron capture reaction cross-sections and astrophysical reaction rates. We consider some of the $(n,\gamma)$ processes consisting a few seed nuclei for the nucleosynthesis in and around the Fe-group for which experimental data are available. We compare our results with those obtained with NLDs from other phenomenological and microscopic models as commonly employed.

\section{Nuclear level densities}

The NLDs are obtained using spectral distribution method \cite{Senkov2013} applied to shell model Hamiltonian with a realistic residual interaction. Within the spectral distribution method, one calculates first and second moments of Hamiltonian for the full configurational space. These moments are then used to construct the Gaussian distribution of the levels which eventually give rise to the level density. For a given isotope $(Z,N)$, the valence nucleons can be distributed in many ways over available orbitals. Each of these configurations is known as
partition, $p$ which contains $D_{\alpha p}$ many-body states with exact quantum numbers $\alpha$. The states present in a given partition are distributed over some energy region as a result of interactions inside the partition. For each partition, the statistical average of an operator $\hat{O}$ over the states is defined through the corresponding
trace,
\begin{eqnarray}
\langle \hat{O} \rangle = \frac{1}{D_{\alpha p}} Tr^{\alpha p} \hat{O}
\end{eqnarray}

\begin{table*}[!htb]
\caption{\label{tab:egs} The ground state energies $E_{g.s.}$ (MeV) for different nuclei required to compute the cross-sections for the $(n,\gamma)$ processes considered. The $E_{g.s.}$ are obtained using shell model Hamiltonian with GXPF1A residual interaction. The exponential convergence \cite{Horoi2002} is used for the cases where the full JT-space dimension $(N_{full})$ exceeds $\sim 5 \times {10}^6$.  The exponential converged value for $^{51}$V is -125.46 MeV which is nearly equal to -125.54 MeV obtained for full model space calculation $(E_{g.s.}^{full})$.}
\begin{center}
\resizebox{0.8\textwidth}{!}{%
\begin{tabular}{c | c c c | c c c | c c c }
\hline
\multicolumn{1}{c|}{}&
\multicolumn{3}{c|}{$^{50}$V$(n,\gamma)^{51}$V}&
\multicolumn{3}{c|}{$^{54}$Fe$(n,\gamma)^{55}$Fe}&
\multicolumn{3}{c}{$^{58}$Ni$(n,\gamma)^{59}$Ni}\\
\hline
Channels& Nucleus & $N_{full}$ & $E_{g.s.}$ & Nucleus & $N_{full}$ & $E_{g.s.}$& Nucleus & $N_{full}$ & $E_{g.s.}$\\
\hline
$(n,\gamma)$& $^{51}$V & 1242538 & -125.54 & $^{55}$Fe & 25743302 & -184.76 & $^{59}$Ni & 76528736 & -236.37 \\
$(n,n)$     & $^{50}$V & 795219 & -114.79  & $^{54}$Fe & 5220621 & -175.73  & $^{58}$Ni & 21977271 & -227.59 \\
$(n,p)$     & $^{50}$Ti & 39899 & -109.86  & $^{54}$Mn & 17069465 & -167.40 & $^{58}$Co & 37534140 & -218.90 \\
$(n,2n)$    & $^{49}$V & 422870 & -105.63  & $^{53}$Fe & 21131892 & -162.56 & $^{57}$Ni & 76528736 & -215.55 \\
$(n,np)$    & $^{49}$Ti & 150632 & -99.24  & $^{53}$Mn & 14123745 & -158.51 & $^{57}$Co & 90369789 & -210.34 \\
$(n,2p)$    & $^{49}$Sc &	28603 & -90.41  & $^{53}$Cr & 3776746 & -151.70  & $^{57}$Fe & 13436903 & -203.30 \\
\hline
\end{tabular}}
\end{center}
\end{table*}

In particular, the  centroid energy of the partition is the first moment of the Hamiltonian,
\begin{eqnarray}
E_{\alpha p} = \frac{1}{D_{\alpha p}} Tr^{\alpha p} \hat{H}
\end{eqnarray}
This comes directly from the diagonal elements of the Hamiltonian matrix. The second moment of the Hamiltonian,
\begin{eqnarray}
\sigma^2_{\alpha p} = \langle H^2 \rangle_{\alpha p} -E^2_{\alpha p} = \frac{1}{D_{\alpha p}} Tr^{\alpha p} H^2 -E_{\alpha p}^2
\end{eqnarray}
is determined by the off-diagonal elements of the Hamiltonian matrix including the interaction between partitions. No matrix diagonalization is required as this quantity can be read directly from the Hamiltonian matrix. The actual distributions are close to the Gaussians which is a manifestation of quantum complexity and chaotization \cite{Mon1975,Brody,Wong,Kota2010}. Finally, the level density $\rho(E; \alpha)$ is found by summing the Gaussians weighted over their dimensions for all partitions at given energy $E$ and with quantum numbers $\alpha$,
\begin{eqnarray}
\rho(E; \alpha)=\sum_p D_{\alpha p} G_{\alpha p}(E)
\end{eqnarray}
The best results are obtained by using finite range or truncated Gaussians,
\begin{eqnarray}
G_{\alpha p}(E)= G(E - E_{\alpha p} + E_{\rm g.s.}; \sigma_{\alpha p}) \label{egs}
\end{eqnarray}
for each partition.
\begin{figure}[!htb]
\vspace{-0.3cm}
\begin{center}
\includegraphics[width=0.45\textwidth]{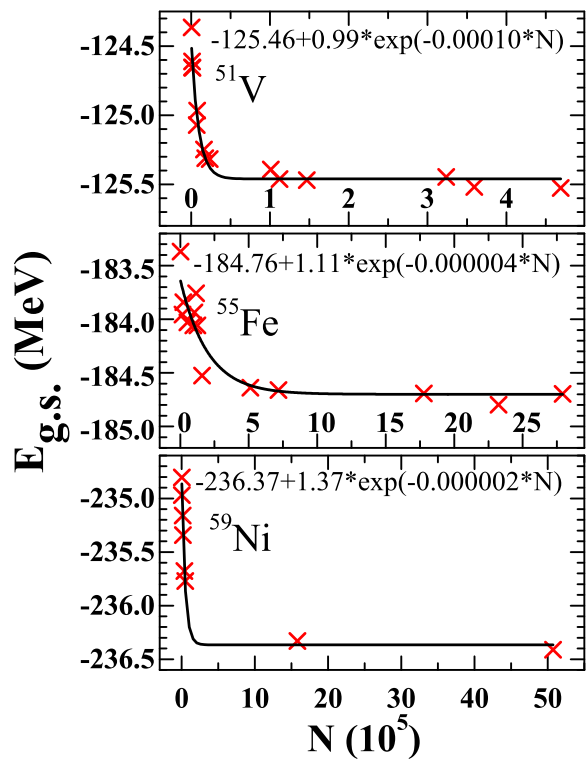}
\end{center}
\vspace{-0.5cm}
\caption{Shell model ground state energies for $^{51}$V, $^{55}$Fe and $^{59}$Ni corresponding to different truncations with dimensions $N$ (red crosses). These energies are fitted to $a+b e^{-cN}$ as indicated by solid line. The exponential converged ground state energies for full $pf$-model space are obtained using this fitted expression with $N=N_{full}$.}\label{fig:egs}
\end{figure}
Removing unphysical tails, the Gaussians are cut off at a distance $\sim 2.6 \sigma_{\alpha p}$ from the corresponding centroid and then re-normalized \cite{Senkov2013}. The ground state energies $E_{\rm g.s.}$ appearing in Eq. (\ref{egs}) must be calculated using full Hamiltonian matrix in order to be consistent with the first and second moments. Practically, it is convenient to calculate the invariant traces in the M-scheme. When $\rho(E;M)$ for a given parity is computed, the level density $\rho(E;J)$ for certain spin $J$ is found as the difference of $\rho(M=J)$ and $\rho(M=J+1)$.
\par

To illustrate, we have considered the reaction cross-sections for $^{50}$V$(n,\gamma)^{51}$V,
$^{54}$Fe$(n,\gamma)^{55}$Fe, and $^{58}$Ni$(n,\gamma)^{59}$Ni processes. In addition to emission of $\gamma$, contributions from ejection of neutron $(n)$ and proton $(p)$ are also included. This leads to the possible reaction channels namely, $(n,\gamma)$, $(n,n)$, $(n,p)$, $(n,2n)$, $(n,np)$, and $(n,2p)$ which require the computation of NLDs for 18 nuclei, 6 for each process. For these nuclei, we compute NLDs with $pf$-model space assuming $^{40}$Ca as a core. We use
GXPF1A residual interaction \cite{Honma2004} which is well known to reproduce the binding energies, electromagnetic moments and transitions as well as excitation spectra for $pf$-shell nuclei.\par

\begin{table*}[!htb]
\caption{\label{tab:resonance} The values of s-wave neutron resonances $(D_{0})$ are calculated for $pf$-model space considering parity equilibration scheme (given by Eq. (\ref{Eq:parity})) at $E_{0}$=$S_n$ and 0.8$S_n$, labelled by SDM and SDM*. These results are compared with the experimental data along with other microscopic mean-field models, HFB-u (un-normalized) and HFB as well as the phenomenological models, BSFG and GSM. The neutron separation energy $S_{n}$ for the product nuclei and the angular momentum $J_{t}^{{\pi}_t}$ for the target nuclei are listed.}
\resizebox{0.65\textwidth}{!}{%
\begin{tabular}{c | c  c | r@{\hskip 0.1in} c @{\hskip 0.1in}c @{\hskip 0.1in}c@{\hskip 0.1in} c@{\hskip 0.1in} c@{\hskip 0.1in} c@{\hskip 0.1in} c}
\hline
\multicolumn{1}{c|}{Nucleus}&
\multicolumn{1}{c}{$J_t^{{\pi}_{t}}$}&
\multicolumn{1}{c|}{$S_{n}$}&
\multicolumn{8}{c}{$D_0$ (\text{keV})}\\
\cline{4-10}
 &  & (MeV) & Exp.&HFB-u&HFB&BSFG&GSM& SDM&SDM*&
\\
\hline
$^{49}$Ti & $0^{+}$           & 8.142  & $18.3\pm2.9$ &24.6&18.6&19.1&15.2&174.4&104.9
\\
$^{50}$Ti & $\frac{7}{2}^{-}$ & 10.939 & $4\pm0.8$    &1.1&4&4.1&5.2&19.6&10.8
\\
$^{51}$V  & $6^{+}$           & 11.051 &$2.3\pm0.6$    &0.9&1.5&2.8&4.8&6.2&3.5
\\
$^{53}$Cr & $0^{+}$           & 7.939  &$43.4\pm4.4$ &30.1&47.7&43&79&66.2&38.4
\\
$^{55}$Fe & $0^{+}$           & 9.298  &$18\pm2.4$     &5.5&15.9&16.3&104&31.7&18.4
\\
$^{57}$Fe & $0^{+}$           & 7.646  &$25.4\pm2.2$   &22&26.4&24.5&54.2&34.8&21.2 \\
$^{59}$Ni & $0^{+}$           & 8.999  &$13.4\pm0.9$   &6.1&12.8&14.1&95.7&23.8&13.9\\
\hline
\end{tabular}}
\end{table*}

It may be pointed out that even a smaller inaccuracy in $E_{\rm g.s.}$ (of the order of 0.5 MeV) would cause large uncertainties in NLDs ($\sim$ 30-20$\%$ for excitation energies $\sim$ 5-10 MeV) that can significantly affect the reaction cross-sections and astrophysical reaction rates.
The accurate values of $E_{\rm g.s.}$ for the case of $pf$-model space are calculated using NushellX@MSU \cite{BROWN2014115}. The calculation of $E_{\rm g.s.}$ turns out to be cumbersome in few cases due to large dimension corresponding to full model space $(N_{full})$ of the Hamiltonian matrix. In such cases, we recourse to the exponential convergence method \cite{Horoi2002}. The ground state energies obtained for several restricted model spaces are fitted to the exponential function of the form $a+b e^{-cN}$. Once the parameters $a$, $b$ and $c$ are known, the exponentially converged value of $E_{\rm g.s.}$ for $pf$-model space can be obtained with $N=N_{full}$. For instance, in Fig. \ref{fig:egs}, the values of $E_{\rm g.s.}$ with different truncations are plotted for $^{51}$V, $^{55}$Fe and $^{59}$Ni. We have also performed full model space calculations for $E_{\rm g.s.}$ with full $JT$-space dimension $(N_{full})$  $\lesssim 5 \times {10}^6$. The exponentially converged values for $E_{\rm g.s.}$ are found to be accurate within 0.1 MeV. In Table \ref{tab:egs}, we list the values of $E_{\rm g.s.}$ along with $N_{full}$ for $pf$-model space for all the 18 nuclei relevant to the present work. Once the ground state energies are known, the NLDs are calculated within the SDM using MM code \cite{Senkov2013}. \par

\begin{figure}[!htb]
\begin{center}
  \includegraphics[width=0.45\textwidth]{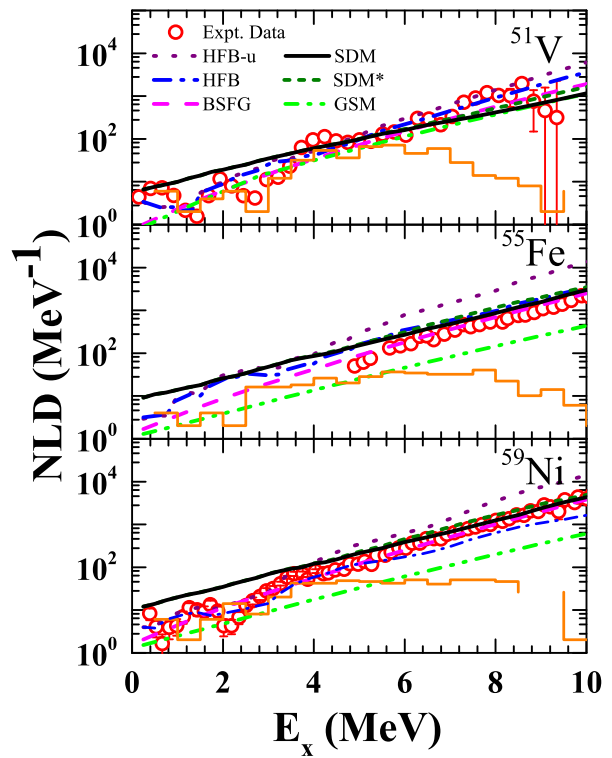}
\end{center}
\caption{Nuclear level densities as a function of excitation energies (E$_\text{x}$) for $^{51}$V, $^{55}$Fe and $^{59}$Ni nuclei obtained from SDM in $pf$-model space by considering parity equilibration scheme (given by Eq. (\ref{Eq:parity})) at $E_{0}$=$S_n$ and 0.8$S_n$ labelled as SDM and SDM*, respectively. These results are compared with the experimental data \cite{Larsen2006,ramirez2015,Voinov2012} along with other microscopic mean-field models, HFB-u (un-normalized) and HFB as well as the phenomenological models, BSFG and GSM. Histograms represent the NLDs obtained from low-lying discrete levels \cite{Capote2009}.}
\label{nld}
\end{figure}


The NLDs for both the parities would require the model space consisting of single-particle states with different parities. The $pf$-model space can yield the NLDs only for  a single parity for a given nucleus. In even-A nuclei, it will correspond to only positive parity states and those for odd-A nuclei to the negative parity states. Recently, Ormand and Brown \cite{Ormand2020} have estimated the values of s-wave resonance spacing $D_0$ for $^{57}$Fe using the NLDs for ${1/2}^-$ states, instead of ${1/2}^+$ states, obtained within the $pf$-model space. The values of $D_0$ so determined agreed reasonably with the experimental data. It was thus inferred that the parity equilibration for $^{57}$Fe occurs near neutron separation energy ($S_n$). The NLDs for $J^\pi = 2^+ $ and $2^-$ states extracted from the measurements of quadrupole giant resonances in $^{58}$Ni \cite{kalmykov2007} showed that parity equilibration takes place at lower energy than the predictions of SMMC calculations for $pfg_{9/2}$ model space \cite{nakada1997,nakada1998}. Similar conclusions were also drawn on the basis of calculations performed for sufficiently large model space \cite{mocelj2007,houcke2009}.
A simple formula \cite{alhassid2000} further suggested the parity equilibration to occur exponentially with increase in temperature.
We construct the level density for positive (negative) parity for odd (even) nucleus through the parity equilibration. We use a simple scheme for the parity equilibration such as
\begin{equation}
    \frac{\rho_{+} (E_\text{x})} {\rho_{-} (E_\text{x})} = \frac{1}{1+e^{-\beta(E_\text{x}-E_{0})}}
    \label{Eq:parity}
\end{equation}
where $\rho_{+} (E_\text{x})$ and $\rho_{-} (E_\text{x})$ are the level densities of positive and negative parities at excitation energy $E_\text{x}$ for odd A nuclei. The $\rho_{+}$ and $\rho_{-}$ will be reversed for the case of even A nuclei. The parameters $\beta$ and $E_{0}$ can be adjusted to obtain required excitation energy dependence of the parity equilibration. For simplicity, we consider $\beta$=1. We
obtain the NLDs using SDM with $pf$-model space by considering above parity equilibration scheme for $E_{0}$=$S_n$ and 0.8$S_n$, labelled by SDM and SDM*, respectively. These NLDs are employed to calculate the  $D_0$, cross-sections and reaction rates.\par

In Fig. \ref{nld}, we show the shell model NLDs, labelled as SDM and SDM*, for $^{51}$V, $^{55}$Fe and $^{59}$Ni nuclei as representative examples and compare them with the existing experimental data \cite{Larsen2006, ramirez2015, Voinov2012}.
 The NLDs obtained from the low-lying experimental discrete levels \cite{Capote2009} are also shown for comparison. Our calculated NLDs are in an overall agreement with the experimental data for all the three nuclei. For the comparison, we also display the NLDs obtained from Hartree-Fock-Bogoliubov (HFB) approach for Skyrme type effective force BSk14 \cite{Goriely2008}. These NLDs are obtained using combinatorial method at lower excitation energies and using statistical method at higher excitation energies. The HFB results are further normalized to the experimental data at low energies and the neutron separation energy \cite{Goriely2008}. We also display un-normalized HFB results as labelled by HFB-u. The NLDs for HFB-u deviate noticeably for $^{55}$Fe and $^{59}$Ni nuclei. The curves marked by BSFG and GSM correspond to phenomenological Back-shifted Fermi gas model \cite{Dilg1973} and Generalized Superfluid model \cite{Ignatyuk1979,Ignatyuk1993}, respectively.\par

\begin{figure*}[!htb]
\vspace{-0.8cm}
\begin{center}
\includegraphics[width=6in, height=4in]{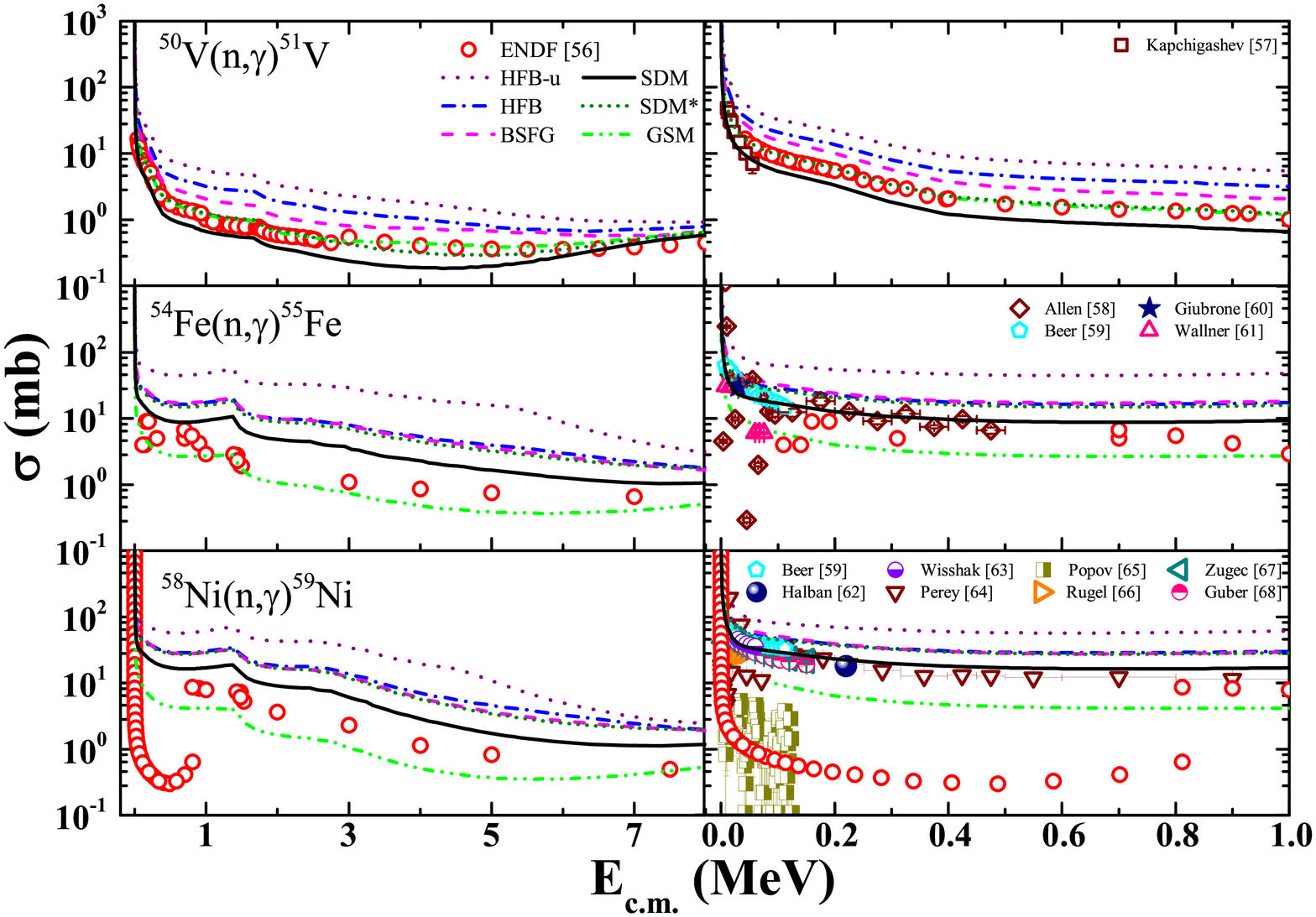}
\end{center}
\vspace{-0.5cm}
\caption{Cross-sections for $(n,\gamma)$ processes as a function of incident neutron energies in center-of-mass frame $(E_\text{c.m.})$ using NLDs from SDM and SDM*  compared with those obtained from other models (as shown in Fig. \ref{nld}), see text for details. Evaluated data are adopted from ENDF. The cross-sections at low $E_\text{c.m.}$ are shown in the right panels together with the respective experimental data.}\label{xs}
\end{figure*}


The cross-sections and reaction rates, relevant to the astrophysical process are predominantly governed by the NLDs corresponding to spin and parity which determines the level spacings $D_0$ for s-wave neutron resonance.
The $D_0$ can be evaluated as \cite{Goriely2008}
\begin{eqnarray}
D_0 =\left\{\begin{array}{cc}
    \frac{1}{\rho(S_n, J_t + 1/2, \pi_t)+ \rho(S_n, J_t - 1/2, \pi_t)} & \text{for} \quad J_t \neq 0,   \\
  & \\
     \frac{1}{\rho(S_n, 1/2, \pi_t)} & \text{for} \quad J_t = 0
     \end{array} \right .
\label{D0}
\end{eqnarray}
where $J_{t}$ and $\pi_t$ are the spin and parity of target nucleus, $S_{n}$ is the neutron separation energy for the product nucleus. It is evident from the Eq. (\ref{D0}) that the calculation of $D_{0}$ for even-even target nucleus requires the level density for ${1/2}^{+}$ in the product nucleus. Similarly, if the target is not even-even then the level densities for the product nucleus are required for the spins J$_{t}\pm1/2$ and parity $\pi_t$. The values of $D_{0}$ are calculated using shell model level densities with $pf$-model space assuming parity
equilibration scheme (Eq. (\ref{Eq:parity})) for $E_{0}$=$S_n$ and 0.8$S_n$ as listed in Table \ref{tab:resonance} along with the available experimental data \cite{Capote2009}. The calculations involve NLDs from ${1/2}^{+}$ states for all the nuclei presented in Table \ref{tab:resonance} except for $^{50}$Ti and $^{51}$V. The values of $D_{0}$ for $^{50}$Ti ($^{51}$V) include the contribution from $3^-$ and $4^-$ (${11/2}^+$ and ${13/2}^+$) spins since $J^\pi_t \neq 0 $ for these cases. The calculated values of $D_{0}$ with $E_{0}$= 0.8$S_n$ (SDM*) are in overall agreement with the corresponding experimental data. The exceptionally larger calculated value of D$_{0}$ for $^{49}$Ti requires further detailed investigation. However, NLDs for this nucleus contribute to $(n, np)$-channel in the neutron capture by $^{50}$V which is not a dominant one. For comparison, we display the results for D$_{0}$ obtained using some selected microscopic mean-field and phenomenological models.\par

\section{$(n,\gamma)$ cross-sections and astrophysical reaction rates}

The $(n,\gamma)$ reaction cross-sections are calculated using TALYS 1.95 computer code \cite{talys1} which takes into account the contributions from three major nuclear reaction mechanisms that include direct reaction, pre-equilibrium emission and compound nucleus. The parameters required for cross-section calculations such as the nuclear masses, discrete levels, decay schemes, OMP, and $\gamma$SF are set to their default values available in TALYS. The extensive database for the NLDs obtained for various phenomenological and microscopic mean-field models are also available in TALYS. These models are the Back-shifted Fermi gas model (BSFG) \cite{Dilg1973}, Gilbert and Cameron model (GC) \cite{Gilbert1965}, Generalized Superfluid model (GSM) \cite{Ignatyuk1979,Ignatyuk1993}, Hartree-Fock for Skyrme type interaction with pairing treated within the Bardeen-Cooper-Schrieffer approximation (HF-BCS) \cite{Demetriou2001}, and Hartree-Fock-Bogoliubov for Skyrme forces (HFB) \cite{Goriely2008}. We compare the reaction cross-sections and astrophysical reaction rates obtained for some of these NLDs with the ones calculated using NLDs from SDM. The cross-sections for $^{50}$V$(n,\gamma)^{51}$V, $^{54}$Fe$(n,\gamma)^{55}$Fe and $^{58}$Ni$(n,\gamma)^{59}$Ni reactions are calculated by including various channels as listed in Table \ref{tab:egs}. Contributions from other channels which include alpha $(\alpha)$, deutron $(d)$, tritium $(t)$ and helium-3 ($^3$He) are found to be insignificant (not shown). \par

In Fig. \ref{xs}, we display our theoretical estimates of $(n,\gamma)$ cross-sections obtained from SDM NLDs corresponding to $E_{0}$=$S_n$ and 0.8$S_n$ (Eq. (\ref{Eq:parity})) and compare them with those obtained for NLDs from other models along with the evaluated data adopted from ENDF database (ENDF/B-VII.1) \cite{endf}. In the left panels, the results are presented for a wide range of incident neutron energy. The corresponding zoomed versions for the lower energies up to $1$ MeV, relevant to astrophysical reaction rates at temperatures upto the order of GK, are presented in the right panels. The cross-sections based on the NLDs from SDM are in an overall agreement with the evaluated data (left panels) as well as available experimental data from various groups (right panels) \cite{Kapchigashev1965,Allen1976,Beer1975,Giubrone2014,wallner2017,halban1938,Wisshak1984,Perey1993,popov2000,Rugel2007,Zugec2014,Guber2010}.
For the low incident neutron energies, our SDM results are in agreement with the experimental data which emphasizes the role of residual interaction in the astrophysical regime. Also, the results from HFB, BSFG and GSM are found to be in a reasonable agreement with the experimental cross-sections as the NLDs for these models are appropriately normalized with the measured ones. However, the results obtained by HFB-u, which correspond to un-normalized NLDs, show noticeable deviations from the measured cross-sections for neutron capture reactions. \par
\begin{figure}[!htb]
\vspace{-0.0cm}
\begin{center}
\resizebox{0.45\textwidth}{!}{%
  \includegraphics{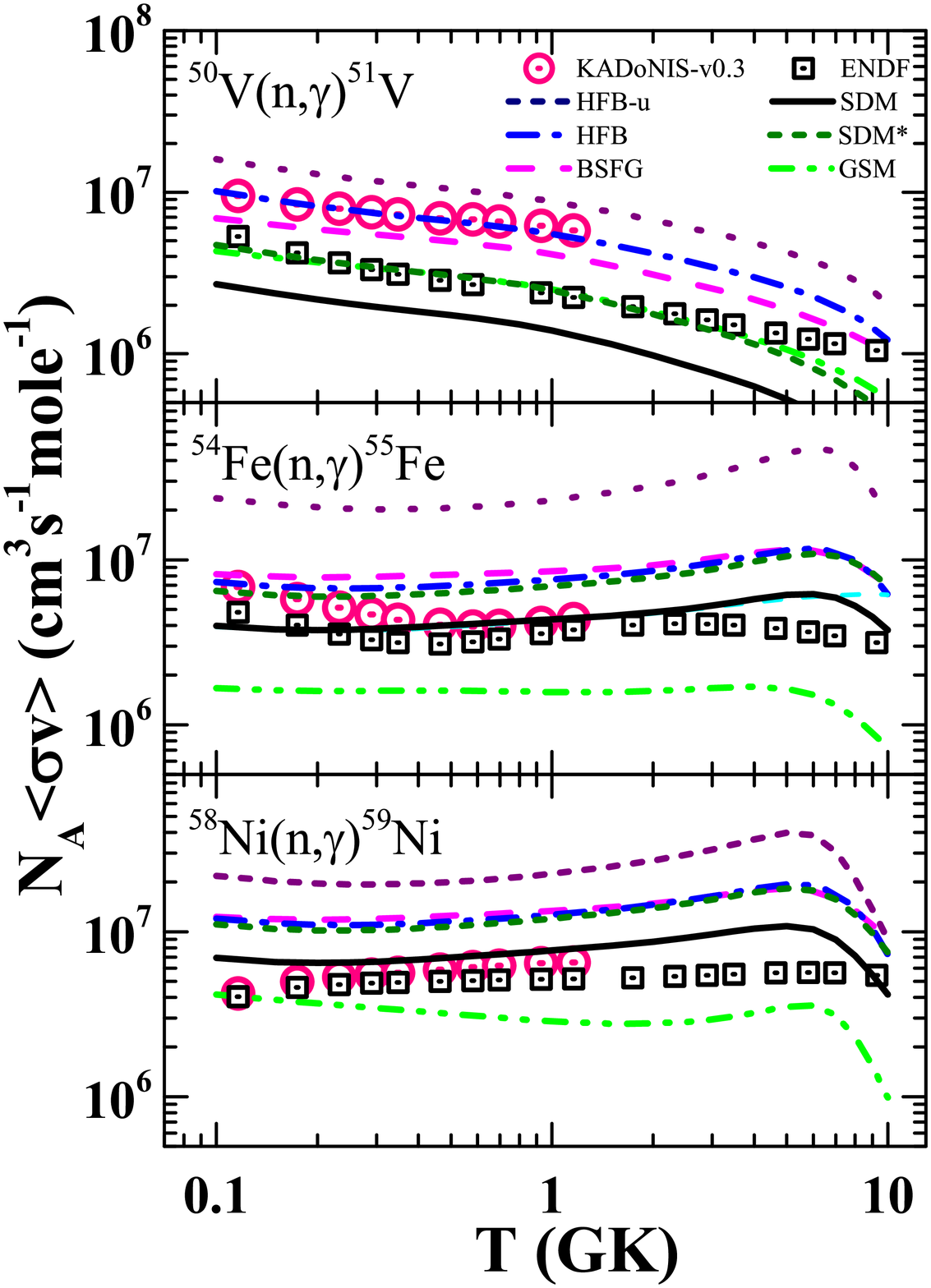}
}
\end{center}
\vspace{-0.5cm}
\caption{The astrophysical reaction rates as a function of temperature using NLDs from SDM and SDM* compared with those obtained for other models (as shown in Fig. \ref{nld}). For comparison, the recommended values from ENDF \cite{endf} and ‘KADoNiS v0.3’ \cite{Dillmann2006} are shown. For the case of $^{50}$V$(n,\gamma)^{51}$V, KADoNiS estimates are purely theoretical.}\label{fig:rr}
\end{figure}

For the reaction network, one needs astrophysical reaction rates as one of the inputs. Once the variation of cross-section with energy is known, the astrophysical reaction rates $N_{A}<\sigma\text{v}>$ can be computed with $N_{A}$ being the Avogadro number and $<\sigma\text{v}>$ is the Maxwellian average, where $\text{v}$ is the relative velocity of neutron. These averages are computed at a fix temperature which ranges from 0.1 GK to 10 GK. For these temperatures, cross-sections at energies within the range of few keV to $\sim$1 MeV contribute maximally. We show in Fig. \ref{fig:rr}, the astrophysical reaction rates obtained by using the NLDs from SDM and compare them with the recommended values from ENDF \cite{endf} and ‘KADoNiS v0.3’ \cite{Dillmann2006}. These results are basically the Maxwellian average of cross-sections shown in Fig. \ref{xs}. The results from HFB, BSFG and GSM are also shown for comparison. SDM results with 0.8$S_n$$\leq$ $E_{0}$$\leq$ $S_n$ explain quite well the recommended values from ENDF \cite{endf} in all the cases. \\
\section{SUMMARY and OUTLOOK}
We obtain realistic NLDs within the framework of spectral distribution method applied to many-body shell model Hamiltonian for $pf$-model space. A particular attention has been paid to calculate the accurate ground state energy since it is a crucial input in the SDM calculations. To incorporate the NLDs of opposite parities in $pf$-model space, an appropriate parity equilibration scheme has been used. The NLDs so obtained and s-wave neutron resonance spacings agree reasonably well with the available experimental data. We further compute reaction cross-sections and astrophysical reaction rates for the neutron capture processes such as $^{50}$V$(n,\gamma)^{51}$V, $^{54}$Fe$(n,\gamma)^{55}$Fe and $^{58}$Ni$(n,\gamma)^{59}$Ni. The calculated reaction cross-sections are found to be in harmony with experimental data, particularly for the incident neutron energies of astrophysical interest. Similar is the case for the astrophysical reaction rates, for the temperature ranging from $0.1-10$ GK.\par

Since the present method is quite general and naturally accounts for the collective excitations, therefore, it can be explored in various model spaces and other reactions of astrophysical interest. To obtain the realistic shell-model NLDs of both parities naturally, it would be desirable to perform calculations with larger model spaces which involve huge computation. It would be worthwhile to study the astrophysical reaction rates using shell model NLDs for the neutron-rich nuclei away from the line of $\beta$-stability where the level density may deviate significantly compared to the nearby stable nuclei \cite{AlQuraishi2001, Liddick2016, Roy2020}. Experiments along this direction using inverse kinematics at radioactive ion beam facilities are expected to be operational in near future.

\section*{Acknowledgments}
We are thankful to P. C. Srivastava and R. Senkov for some inputs of the shell model calculations. We also express our gratitude to Pratap Roy for fruitful discussions. TG acknowledges Council of Scientific and Industrial Research (CSIR), Government of India for fellowship Grant No. 09/489(0113)/2019-EMR-I. BM acknowledges the financial support from the Croatian Science Foundation and the \'Ecole Polytechnique F\'ed\'erale de Lausanne, under the project TTP-2018-07-3554 ``Exotic Nuclear Structure and Dynamics", with funds of the Croatian-Swiss Research Programme. She also acknowledges the support received from IIT Ropar, India where a part of this work has been carried out. BKA acknowledges partial support from the Department of Science and Technology, Government of India with grant no. CRG/2021/000101.

\end{document}